\documentclass[12pt,preprint]{aastex}
\input epsf.sty

\newcommand{\nh}{$N_{\rm H}$~}
\newcommand{\nhp}{$N_{\rm H}$}
\newcommand{\sax}{{\it BeppoSAX~}}

\newcommand{\chr}{$\chi^2_r$}
\newcommand{\sgr}{SGR~1900+14}

\begin{document}

\title{Broad band X-ray spectra of short bursts from SGR 1900+14}

\author{M. Feroci, G.A. Caliandro}
\affil{Istituto di Astrofisica Spaziale e Fisica Cosmica, CNR, Roma,
\\
Via Fosso del Cavaliere 100, I-00133 Roma, Italy}
\email{feroci@rm.iasf.cnr.it, caliandr@rm.iasf.cnr.it}

\author{E. Massaro}
\affil{Physics Dept., University of Roma La Sapienza, \\
Piazzale A. Moro 2, I-00185 Roma, Italy \\ IASF, CNR, Sezione di Roma
}
\email{enrico.massaro@uniroma1.it}
\author{S. Mereghetti}
\affil{Istituto di Astrofisica Spaziale e Fisica Cosmica, CNR,
Sezione di Milano, \\ via Bassini 15, I-20133 Milano, Italy}
\email{sandro@mi.iasf.cnr.it}
\author{P.M. Woods}
\affil{Universities Space Researches Association, National Space
Science and Technology Center \\
320 Sparkman Dr. Huntsville, AL 35805 USA}
\email{peter.woods@nsstc.nasa.gov}

\begin{abstract}

We report on the X-ray spectral properties of 10 short bursts from
\sgr~ observed with the Narrow Field Instruments onboard \sax in
the hours following the intermediate flare of 2001 April 18. Burst
durations are typically shorter than 1 s, and often show
significant temporal structure on time scales as short as $\sim$10
ms.  Burst spectra from the MECS and PDS instruments were fit
across an energy range from 1.5 to above 100 keV.  We fit several
spectral models and assumed \nh values smaller than 5$\times
10^{22}$ cm$^{-2}$, as derived from observations in the persistent
emission. Our results show that the widely used \textit{optically
thin thermal bremsstrahlung} law provides acceptable spectral fits
for energies higher than 15 keV, but severely overestimated the
flux at lower energies. Similar behavior had been observed several
years ago in short bursts from SGR 1806-20, suggesting that the
rollover of the spectrum at low energies is a universal property
of this class of sources. Alternative spectral models - such as
two blackbodies or a cut-off power law - provide significantly
better fits to the broad band spectral data, and show that all the
ten bursts have spectra consistent with the same spectral shape.

\end{abstract}

\keywords{stars: neutron -- stars: individual (SGR 1900+14) --
X-rays: bursts }

%\maketitle
%\newpage

\section{Introduction}

The bursting activity from \sgr~ is quite diverse. The most common
bursts observed are the recurrent, short bursts (e.g., Aptekar et
al. 2001). They usually have durations of a few hundreds of ms and
peak luminosities reaching $\sim$10$^{40-41}$ erg s$^{-1}$ (for an
assumed distance of 10 kpc). The most rare but very luminous
events are the so called `giant' flares: to now only one has been
observed from \sgr~ on 1998 August 27 (Feroci et al. 1999, Hurley
et al. 1999b). The burst had a duration of about 300 s and its
time evolution was characterized by a very intense, short and hard
peak followed by a decaying tail that showed coherent pulsations
at the frequency of the persistent pulsed emission. The luminosity
of this `giant' burst was estimated to be greater than 10$^{44}$
erg~s$^{-1}$. Another very bright flare from \sgr~ was observed on
18 April 2001 (Guidorzi et al. 2004), shorter and less intense
than that of 1998, but it too showed periodicity.  Given that both
its duration of $\sim$40 s and peak luminosity of $\sim$10$^{42}$
erg~s$^{-1}$ fell in between the common, recurrent bursts and the
giant flare, this burst was classified as an `intermediate' flare.

Spectral analysis of the SGR bursts has been generally performed
at energies above $\sim$15-20 keV, where the spectra are usually
well represented by an Optically Thin Thermal Bremsstrahlung
(OTTB) law with temperatures ranging from 20 to 40 keV (e.g.,
Aptekar et al. 2001). Detailed spectral studies on uniform samples
of short bursts over wide energy ranges are rare, and we are not
aware of previous studies specifically for \sgr. Laros et al.
(1986) and Fenimore et al. (1994) used the \textit{ICE} data from
5 keV to above 200 keV to study more than 100 bursts from
SGR~1806$-$20. They found that the photon number spectrum of the
detected bursts was remarkably stable despite the large intensity
spread (a factor of 50), and that the low energy (below 15 keV)
data were \textit{inconsistent} with the back extrapolation of an
OTTB model that provided a good fit to the high energy portion of
the spectrum. More recently, Strohmayer and Ibrahim (1998)
confirmed this result using data from \textit{Rossi}XTE.

Qualitatively similar spectral properties were measured during a
bright burst from \sgr\ using {\it HETE-2} (Olive et al.\ 2003).
For this one burst, they noted that the OTTB model overestimated
the flux at low energies ($<$15 keV) and the broad-band spectrum
(7$-$150 keV) was best fit by the sum of two blackbody (BB) laws.

In this paper, we describe the results of a spectral analysis of
ten short bursts from \sgr. \sax pointed at the source $\sim$8
hours after the intermediate flare of April 18 and discovered an
afterglow emission which decayed to the quiescent level in a few
days (Feroci et al. 2003, Woods et al. 2003). During the dimming
afterglow \sgr~ emitted a series of bursts of short duration
($\leq$ 1 s), consistent with the `normal' short bursts. Here we
report on the spectral analysis of these bursts in the energy
range from 1.5 to 100 keV, and show that their spectra are
\textit{inconsistent} with an OTTB law, due to a paucity of
photons in the low energy range.

\section{Observations and Data Analysis}

The \sax observation started on 2001 April 18 at 15:10 UT and
ended on April 19 at 19:38 UT. Net exposures times of the Narrow
Field Instruments (NFIs) were 23.4 ks (LECS), 57.2 ks (MECS) and
41.7 ks (PDS). A point-like source is clearly detected in the LECS
and MECS images. Light curves and spectral data were accumulated
from a circular region with 6$'$ radius, the optimal
signal-to-noise region for which a response matrix is available.

\begin{figure}
\rotatebox{90}{
\epsscale{0.7}
\plotone{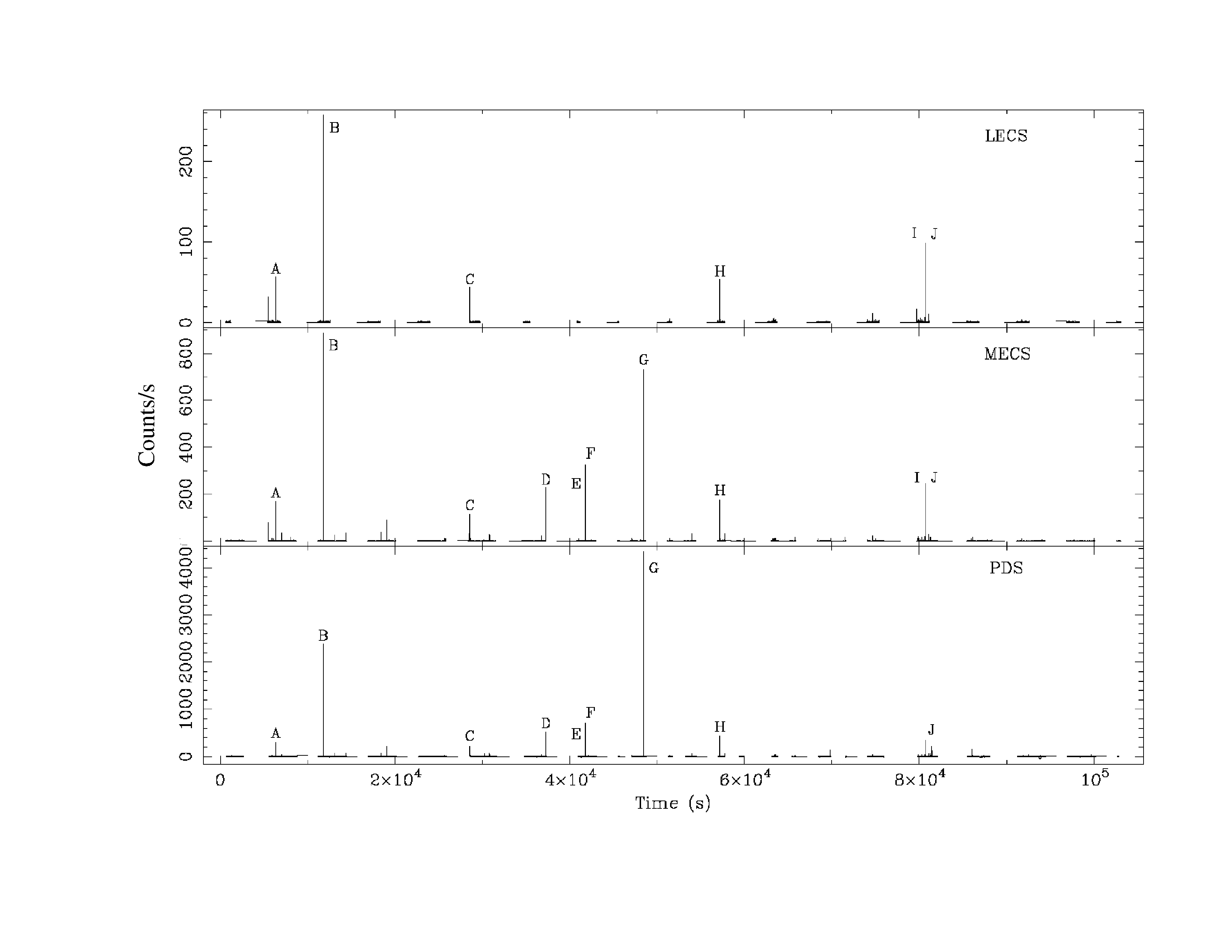}}
\caption{The X-ray light curves of
the afterglow of \sgr~ after the burst of 18 April 2001 in the
LECS (upper panel), MECS (middle panel) and PDS (lower panel)
energy ranges. The ten bursts considered in our spectral analysis
are indicated by the letters from A to J.} \label{lc}
\end{figure}

The MECS (2--10 keV) light curve (Fig. \ref{lc} - middle panel)
shows many short bursts, many having peak count rates more than
two orders of magnitude above the persistent X-ray emission level
from \sgr. In our analysis we considered only the events with an
integral number of counts in the MECS band larger than 100, enough
to obtain significant constraints on spectral fit parameters.
Thus, we selected only ten bursts whose salient properties are
given in Table \ref{burstlist}. We searched the PDS data for
bursts coincident in time with those detected in the MECS light
curve and found significant emission for all ten. Five of the
bursts in Table \ref{burstlist} were also detected in the LECS
data, but the number of counts in the energy range useful for
spectral analysis (0.1-4 keV) is small and does not further
constrain the spectral fit even for the brightest event (burst B).
Thus, we limited our study only to the MECS and PDS data, with an
effective energy range between 1.5 and $\sim$100 keV.

The background counts in the MECS images within an annulus
surrounding the source extraction circle (of equal area), were
found to be less than unity over the bursts time intervals. Hence,
no background subtraction was applied. PDS background spectra were
accumulated from simultaneous data acquired by off-source
detectors. Publicly available matrices were used for all the
instruments. MECS and PDS spectra were binned in order to have a
minimum number of 20 counts in each bin. We used {\sc xspec} 11
for fitting the data. The parameter uncertainties correspond to
$90$\%  confidence for a single parameter, i.e., $\Delta
\chi^2=+2.7$. MECS and PDS spectra were simultaneously fitted
leaving the normalization constant free to vary in the range
0.80--0.90, according to the prescription given by Fiore,
Guainazzi and Grandi (1999).

\begin{figure}
\epsscale{0.8} \plotone{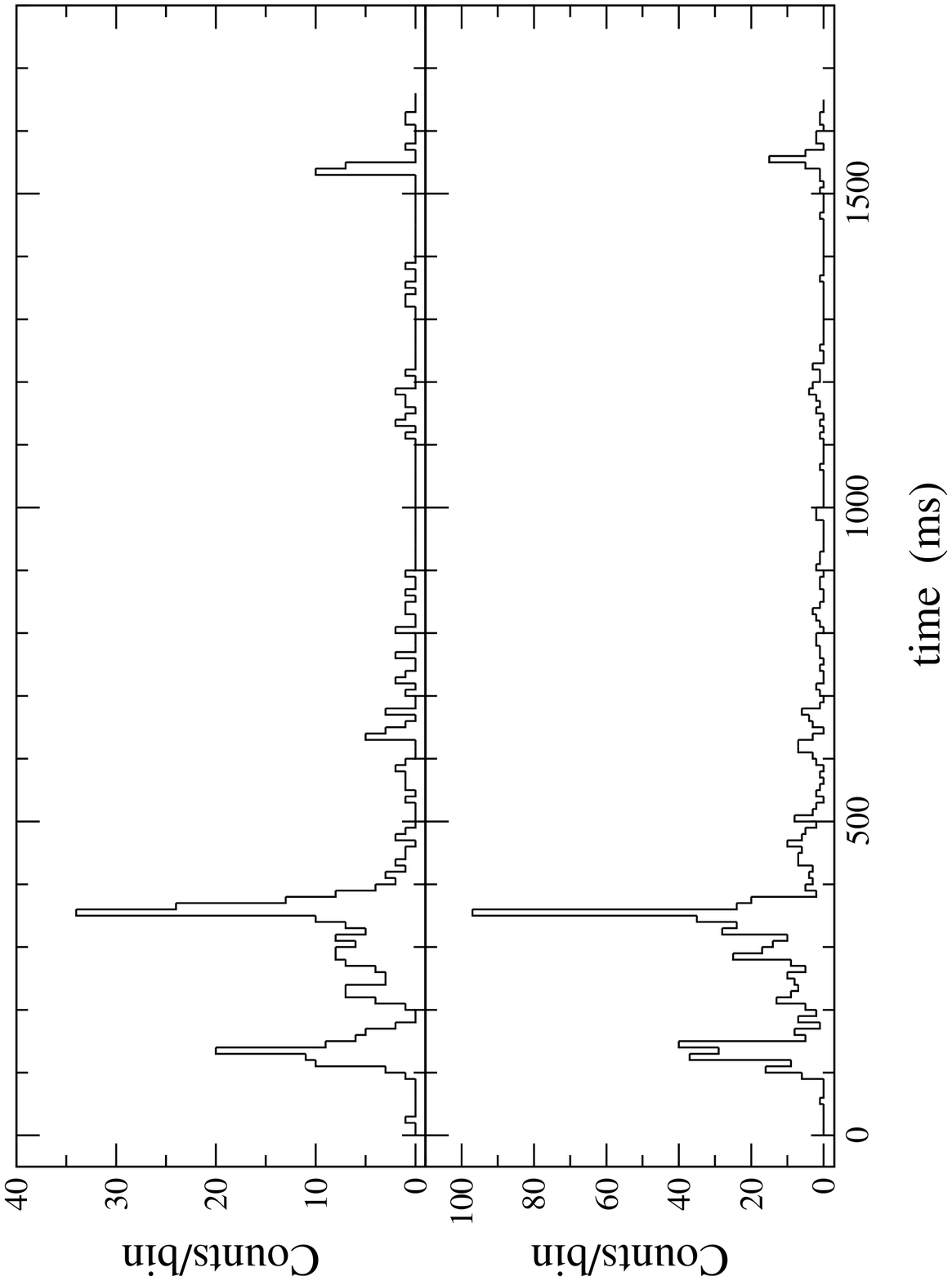} \caption{The X-ray light curves of
the burst D in the MECS (upper panel) and PDS (lower panel) energy
ranges, after correction for the different time off-set (see text
for details). The time bin size is 10 ms. } \label{burst}
\end{figure}

The bursts are often characterized by the presence of several well
separated peaks. When correlating the MECS and PDS light curves,
we detected a systematic time delay of about 50 ms between the two
instruments.  This effect is possibly a consequence of the switch
to the secondary clock onboard the \sax satellite in the year
2000. On the other hand, the similar time structure allowed for a
post-facto empirical time re-alignment of the data, as shown in
Fig. \ref{burst}. Based on this assumption, the time-integrated
energy spectra were extracted for the two instruments on time
intervals encompassing valid simultaneous data from both
instruments. We note that the two brightest bursts, B and G, had
very strong initial spikes with high count rates (larger than
4,000 and 12,000 count/s in the MECS and PDS, respectively) that
saturated the satellite telemetry, effectively truncating the
light curves in the two instruments\footnote{We also checked for
potential dead-time and pile-up effects. For the brightest event,
we estimate a dead-time less than 15\% during the narrowest peak.
The MECS instrument was protected from pulse pile-up through data
processing logic (Boella et al. 1997). Furthermore, we estimate
the pile-up effect in the PDS spectra to be negligible as well
($\sim$1\%).}. We conservatively included in the spectral analysis
only the counts safely away from the saturation edge.

\section{\sax MECS and PDS Spectral Analysis}

A critical parameter in the modeling of these SGR burst spectra is
the column density of neutral absorbing gas along the line of
sight.  The most precise measurement of this parameter comes from
modeling of the persistent X-ray emission.  In quiescence, the
X-ray spectrum of \sgr\ is best fit (Woods et al.\ 1999) by the
sum of a blackbody ($kT \simeq$ 0.5 keV) and a power law (photon
index $\Gamma \simeq$ 1.0-2.5).  For this model, the column
density is consistently measured at N$_{H} \simeq 2.5 \times
10^{22}$ cm$^{-2}$ (Woods et al.\ 1999; Woods et al.\ 2001;
Kouveliotou et al.\ 2001; Feroci et al.\ 2003).  However, this
particular observation where our bursts were recorded took place
during a state of enhanced flux following the 2001 April 18 flare
(Feroci et al.\ 2003), a so-called `afterglow state'.  Fit to a
simple power law, the \nh\ measured at this time is
(4.41$\pm$0.25) $\times$ 10$^{22}$ cm$^{-2}$.  Although not
statistically required by the data, the addition of a blackbody
component to the power law model provides a good fit to the
persistent X-ray emission at this time, with an \nh consistent
with the quiescent state (but more loosely constrained), but with
a temperature close to 1 keV (Mereghetti et al., in preparation).
Indeed, similar variations of the blackbody temperature have been
observed in this and other magnetar candidates (e.g., Lenters et
al. 2003, Woods et al.\ 2004) following intense burst activity.
Thus, it cannot be excluded that the \nh for \sgr~ at the time of
this observation might have been consistent with the quiescent
value of $\sim2.5 \times$ 10$^{22}$ cm$^{-2}$. Here, we will adopt
a conservative approach and allow \nh\ to vary between 2 and 5
$\times 10^{22}$ cm$^{-2}$ for our spectral fits.

\begin{figure}
\centering \epsscale{1.1} \rotatebox{-90}{ \plottwo{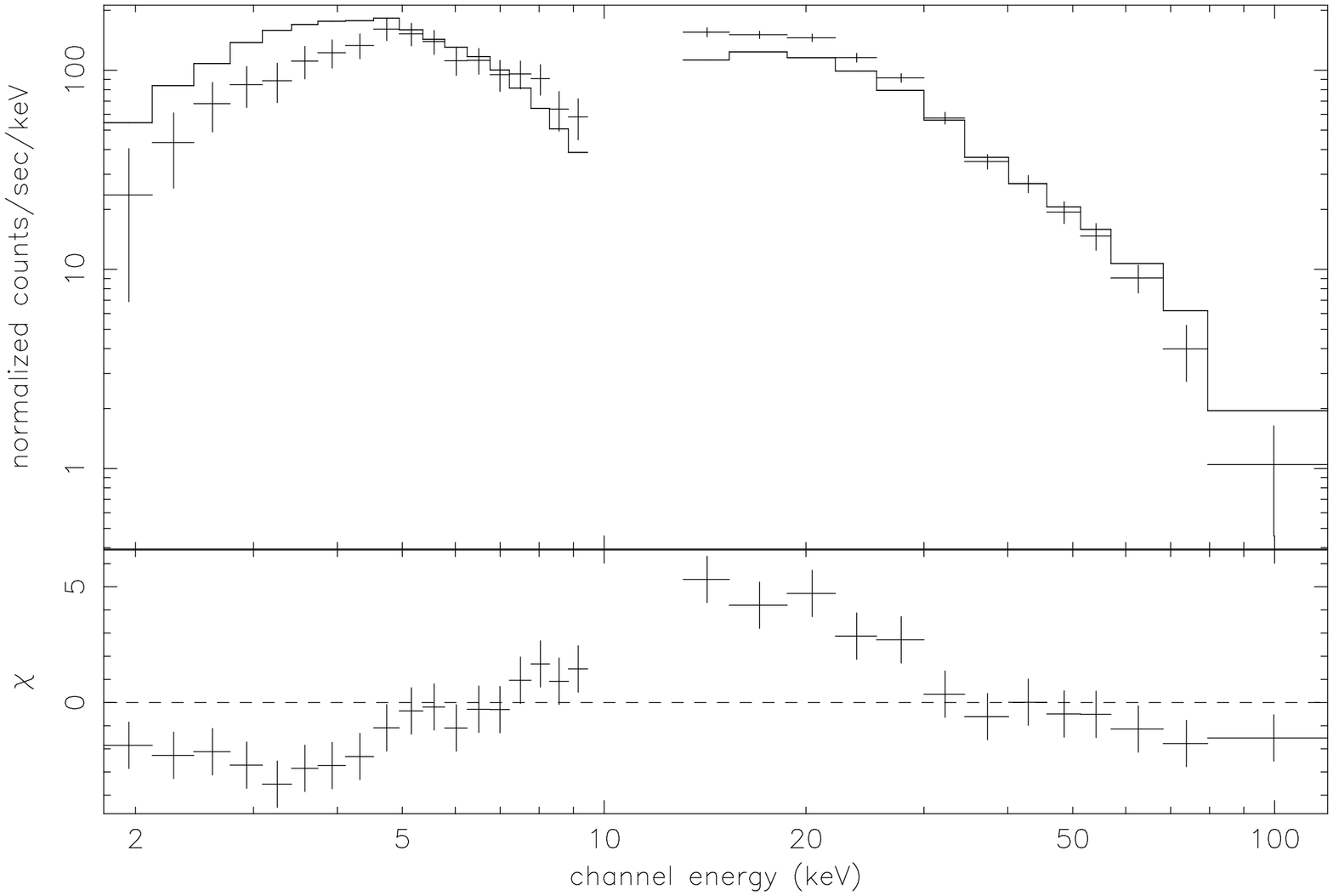}
{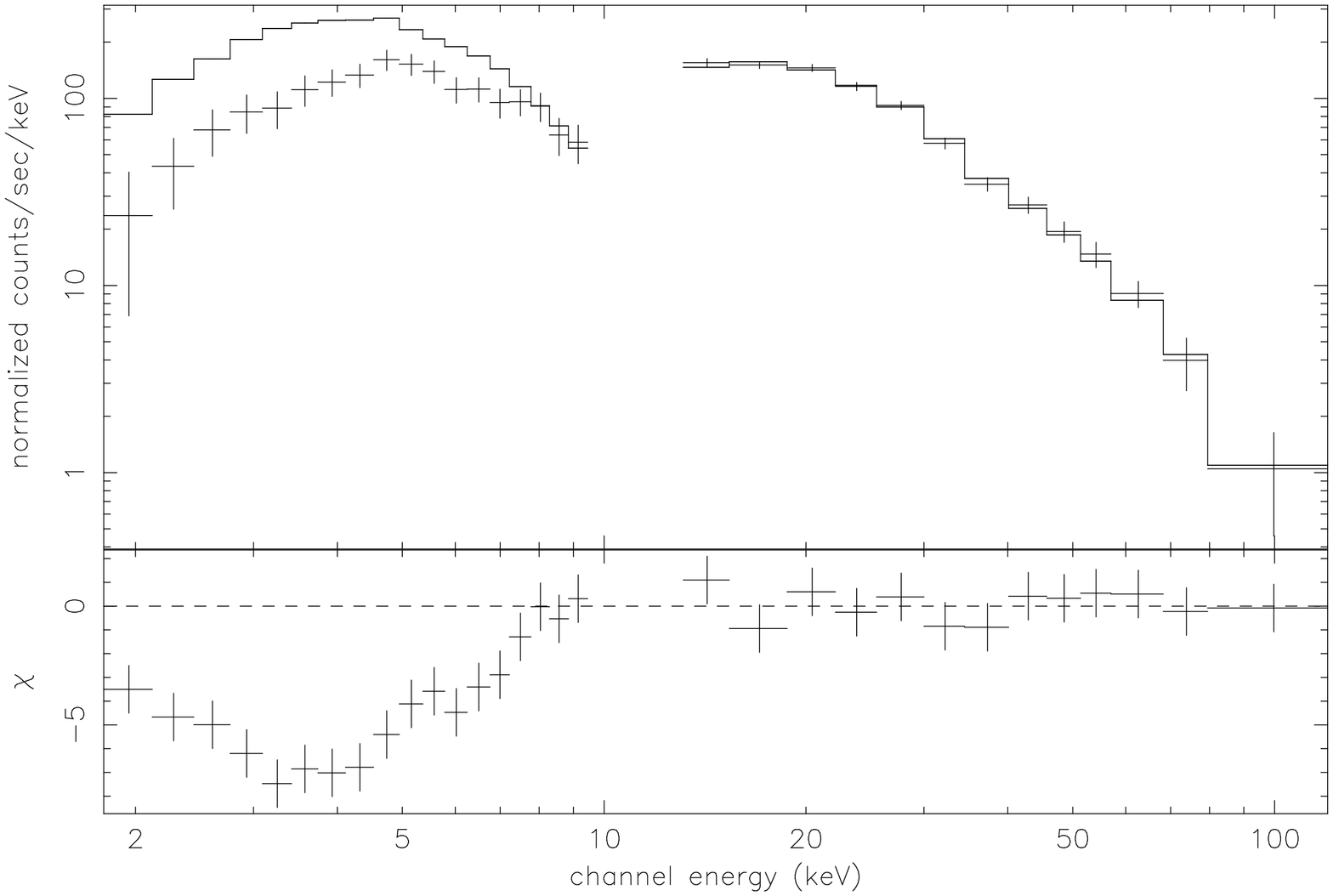}} \caption{The best-fit OTTB model applied to the Average
Spectrum of all the ten bursts, computed from MECS and PDS data
(top panel) and from PDS data only (bottom panel). } \label{fit1}
\end{figure}

One important conclusion from our analysis is that the OTTB model
fails to give acceptable fits to the spectral energy distributions
of all bursts.  Using the combined data set (MECS and PDS) for
each burst and assuming \nh=2.5$\times$10$^{22}$ cm$^{-2}$, the
\chr~ are all in the interval 2--7 (with 12 to 46 degrees of
freedom - d.o.f.), while for \nh=4.4$\times$10$^{22}$ cm$^{-2}$
the \chr~ are lower but remain mostly statistically unacceptable
(\chr~ between 1.4 and 4.8, except for burst J for which
\chr~=1.0, 22 d.o.f.). When we limit the fit to the PDS data only
(E $>$ 15 keV) we generally obtain very good results (0.5$<$ \chr
$<$ 1.2), with the exception of burst C (\chr=2.1). The measured
temperatures ($kT$) are in the range 19--28 keV, in agreement with
the values usually found for the short bursts in this energy range
(e.g., Aptekar et al. 2001). To concisely show the inadequacy of
the OTTB model over the full band, we show a fit to the sum of the
spectra of all bursts (Average Spectrum, Top Panel of Fig.
\ref{fit1}) with \nh fixed at the maximum value within our range,
5$\times$10$^{22}$ cm$^{-2}$ (\chr~ = 4.1 for 30 d.o.f.). If \nh
is left free to vary, a marginally good fit is found (\chr~=1.22
for 29 d.o.f.), but with an unrealistic value of \nh
(13.2$^{+2.2}_{-2.0}$ $\times 10^{22}$ cm$^{-2}$, with
kT=23.1$^{+1.5}_{-1.4}$ keV). A back extrapolation of the fit to
the PDS data only severely overestimates the observed flux at
lower photon energies (Fig. \ref{fit1}, Bottom Panel). We conclude
that the OTTB model fails to fit the broad-band burst spectra of
\sgr.

\begin{figure}\plotone{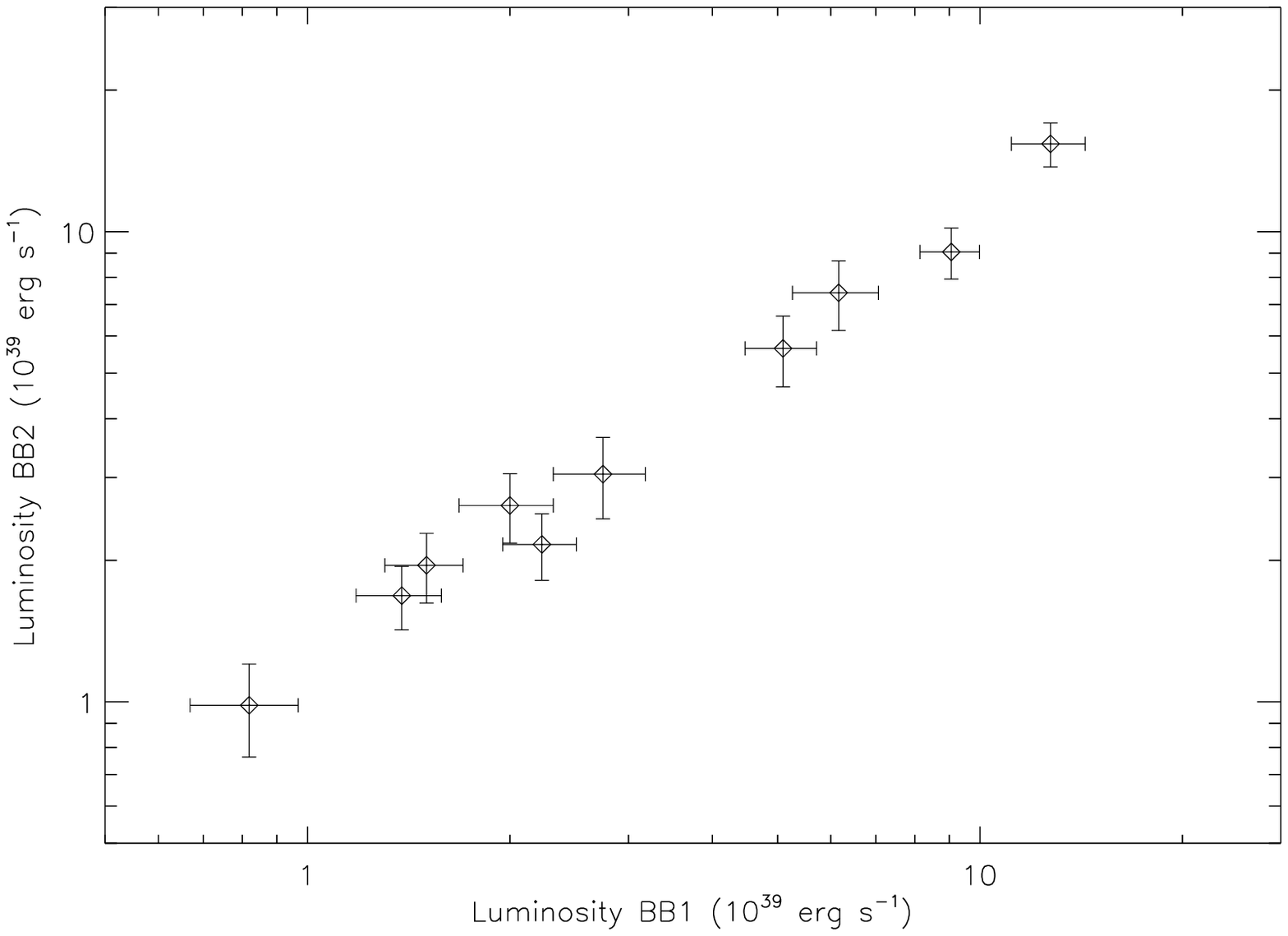}
 \caption{Bolometric luminosity of BB2 (with higher kT) versus
 luminosity of BB1 (with smaller kT) for the ten bursts.}
 \label{corr}
\end{figure}

\begin{figure}
\centering \epsscale{1.1} \rotatebox{-90}{
\plottwo{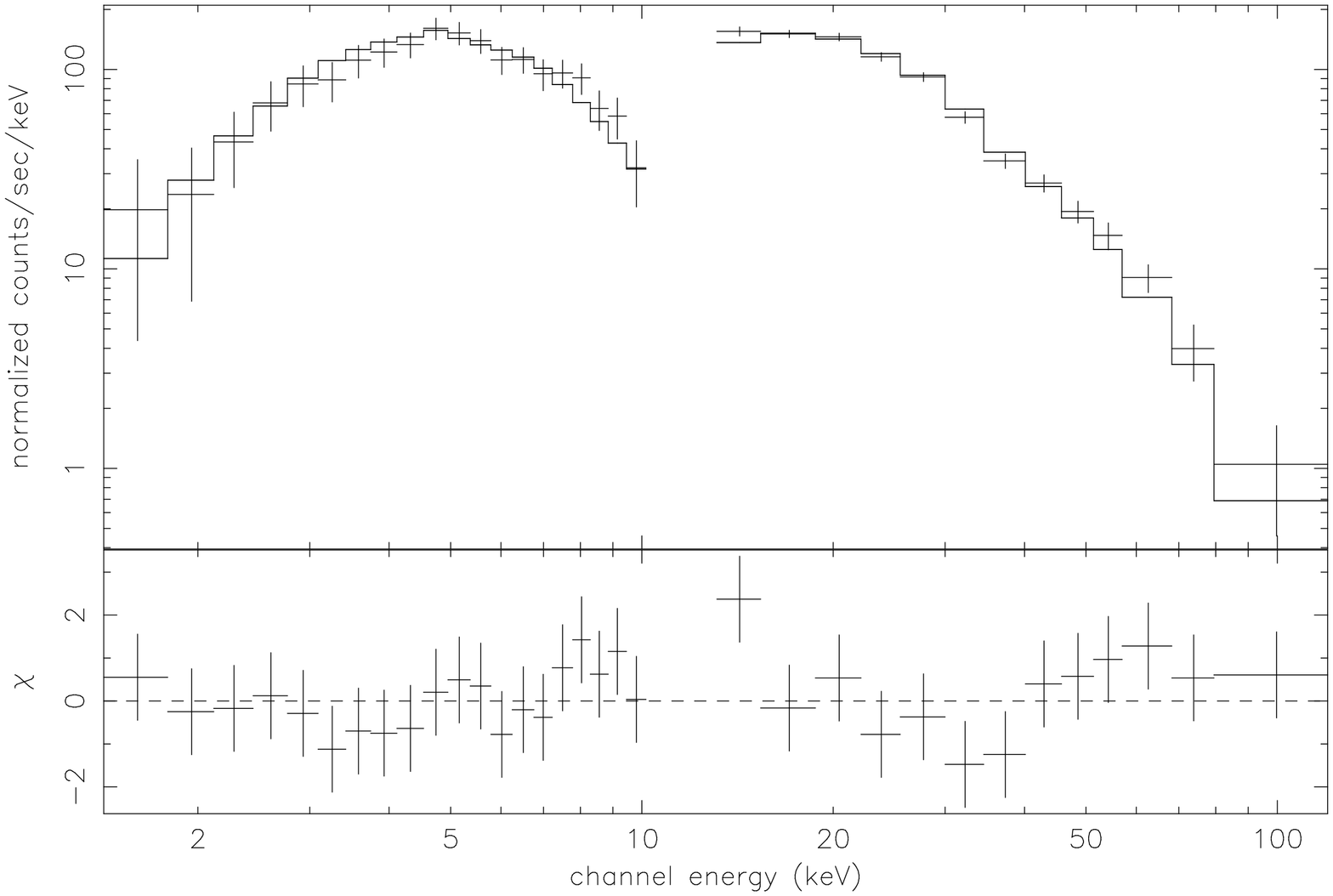}{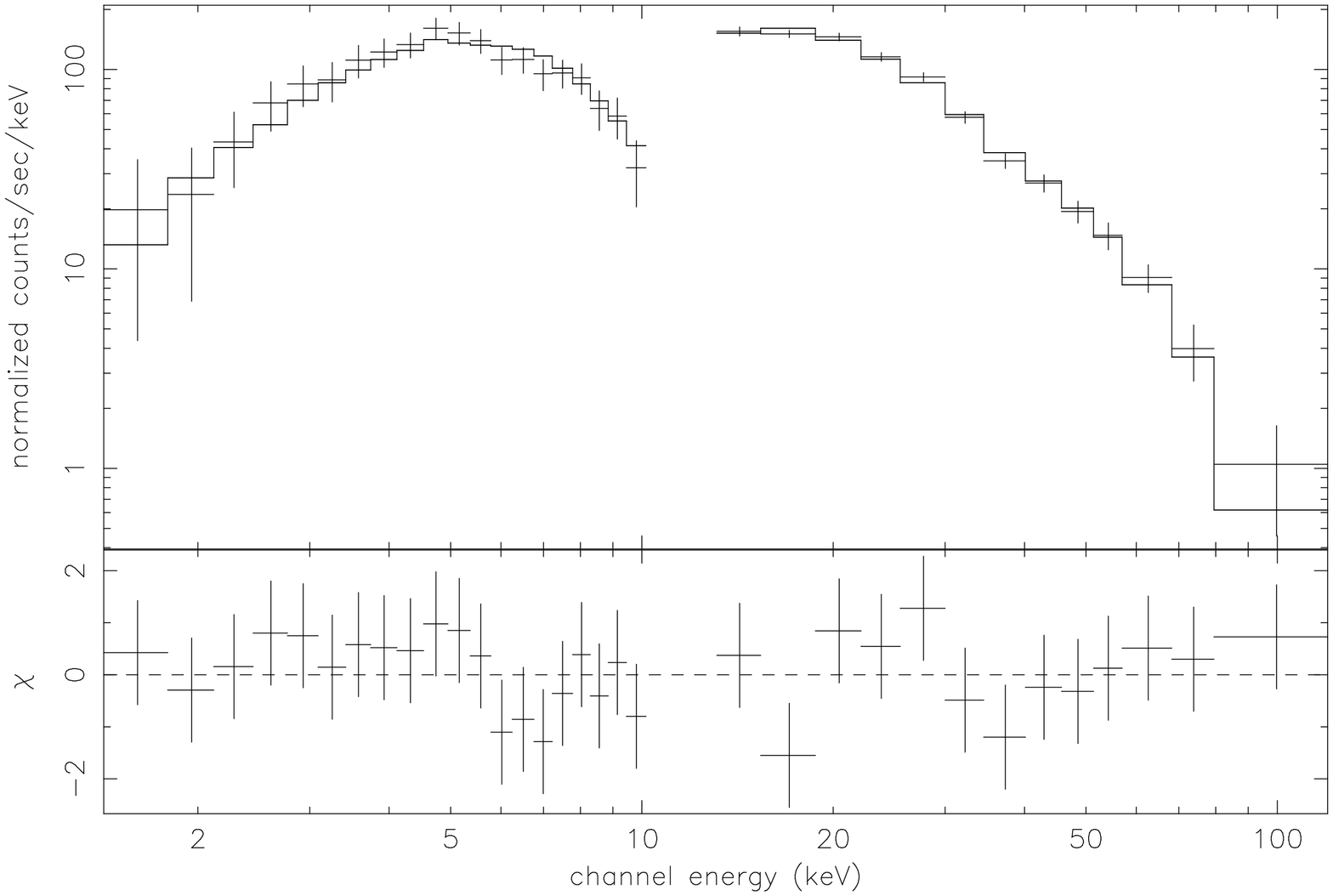}} \caption{The best-fit for the power
law with an exponential cutoff model (upper panel, with \nh
=5$\times 10^{22}$ atoms cm$^{-2}$) and for the sum of two
blackbodies (lower panel, with \nh =2$\times 10^{22}$ atoms
cm$^{-2}$) for the Average Spectrum of all the ten bursts.
}\label{fit2}
\end{figure}

Next, we attempted to find an acceptable model for the burst
spectra that fits both the high and low energies. A single BB
spectrum, which is characterized by a greater curvature, does not
give acceptable best fits for either the individual or the Average
\sax burst spectra (\chr~ $>$ 2.8). Another possible thermal model
is the sum of two BBs, as used by Olive et al. (2003) to fit the
7--150 keV FREGATE spectral data of the 2001 July 2 burst from
this source. When applied either to the individual burst spectra
or to the Average Spectrum it yields a statistically acceptable
\chr. The best fit parameters are given in Table \ref{fit2BB}. The
fit with this model to the Average Spectrum is shown in the upper
panel of Figure \ref{fit2}. The dispersion of the two BB
temperatures is small, consistent within statistics: for the low
temperature BB we found a mean $<kT_1>$ value of 3.23 keV with a
standard deviation of 0.56 keV, while the high temperature
component has a mean $<kT_2>$=9.65 keV and a standard deviation of
0.95 keV. Such a narrow distribution of spectral parameters raises
the question of whether the 10 bursts are all consistent with the
same spectral parameters. We therefore performed a simultaneous
fit to all the burst spectra linking the \nh, $kT_1$ and $kT_2$
between the individual bursts. The fit is satisfactory (line
"Joint Fit" in Table \ref{fit2BB}) with parameters consistent with
those derived from the fit to the Average Spectrum, as well as
with the mean of the parameters derived from the fits to the
individual spectra. Interestingly, the ratio of the bolometric
luminosities of the two BBs is approximately constant for the
different bursts (Table \ref{fit2BB} and Fig. \ref{corr}). We note
also that the best-fit parameters for the spectrum of the 2001
July 2 event obtained by Olive et al. (2003) using the same model
are nicely consistent with our temperature distributions
($kT_{1}$=4.15 keV and $kT_{2}$=10.4 keV).

Finally, we also tested a cut-off power law model. The best-fit
parameters for both individual and Average burst spectra are given
in Table \ref{fitPLC}, for our restricted ranges of \nhp. The fit
results are satisfactory, especially for larger \nhp\ values near
the maximum of our range. As was the case for the two BBs model,
we note that the spectral parameters are rather clustered.  The
cut-off energy distribution is centered at $<E_{c}>$ = 15.8 keV,
with a standard deviation of only 2.3 keV and the mean spectral
index $<\Gamma>$ is 0.4, with a dispersion of 0.2. Also for this
case, a joint fit of all the spectra with linked fit parameters
(except for the normalizations) provided satisfactory results
(line ``Joint Fit" in Table \ref{fitPLC}), consistent with the
results on the Average Spectrum, as well as with the mean of the
values found in the individual fits.

\section{Discussion}

The common, short bursts from \sgr\ are ordinarily detected by the
wide field-of-view experiments typically working above 15-20 keV
or more. This led to a good empirical description of their spectra
using the OTTB model, however, often criticized as non-physical
when applied in this context (e.g., Fenimore et al. 1994). We have
now demonstrated for the short bursts from \sgr~ that the OTTB law
is not acceptable even at an empirical level when the energy
spectrum is studied across a broader energy range. A similar
result was obtained by Fenimore et al. (1994) for SGR 1806-20.

We tested three alternative spectral models, still on
phenomenological grounds, following an approach of simplicity. A
critical spectral parameter for all models is the neutral
absorption column. Leaving \nh\ free, the OTTB model is marginally
acceptable, but the column density required is much larger than
what is found during the persistent X-ray emission seconds before
and after the burst. As a conservative approach, we therefore
decided to constrain in all our fits an equivalent hydrogen column
in the range of what has been measured for the persistent source.
With this assumption in mind, we found that our best-fit models
are two BBs and a cut-off power law. We have no strong statistical
or physical argument to support one or the other. However, we note
that the goodness of the fits with the cut-off power-law model is
more critically sensitive to the choice of \nh. In particular, as
shown in Table \ref{fitPLC}, the fits to the burst spectra
converge to the highest \nh values allowed in the range.  Similar
to the OTTB model, the \chr~ increases as the \nh decreases.
Should the \nh be constrained to small values from future
measurements, this would possibly affect the validity of the
cut-off power law as a spectral model for the bursts.

A striking property of both models is the clustering of the
parameters governing the spectral shape (i.e.\ the temperatures of
the two BB model, and $E_{c}$ and $\Gamma$ for the power law with
an exponential cut-off).  Consistent with previous studies, this
shows the uniformity of the SGR burst spectra with varying burst
intensities. Interestingly, for the two BB model, the luminosity
emitted by the two components is consistent with a constant ratio,
showing that both components scale in intensity for brighter
bursts. As for the cut-off power law model, it has to be noted
that the photon index obtained in all our fits is always
significantly smaller than what is expected for an energy
distribution of particles in thermal equilibrium ($\Gamma$=1).

It is very interesting to note that the short bursts from SGR
1806-20 also possess a similar spectral shape as the ones we
analyzed here from \sgr.  However, Fenimore et al. (1994) measure
an average column density of 1.1$\times$10$^{24}$ cm$^{-2}$ using
an OTTB model with $kT\sim$22 keV.  This column density is almost
an order of magnitude larger than the value we measure for \sgr\
using the same model, with similar $kT$. Thus, it appears that the
roll-over at low energies in SGR~1806$-$20 is more severe than it
is for \sgr.

Although perhaps quantitatively different, the spectral
similarities for \sgr~ and SGR 1806-20 may reflect a
\emph{universal} emission mechanism for short bursts independent
of the specific source. This assertion would need support from
similar analysis of short bursts from the other SGR sources. It is
worth noting that such a universality is already suggested in the
mechanism for the emission of the giant flares (e.g., Mazets et
al. 1999, Feroci et al. 2001) for the two cases known so far from
SGR 0526-66 and SGR 1900+14.

Two conclusions that we can draw from our observations are:
\textit{i}) the spectral shape of the individual bursts remains
stable independent of the brightness, duration and light curve
complexity of the events; \textit{ii}) although we have identified
two spectral models that provide a good description of the data,
it remains to be seen whether these models truly reflect the
physics behind the emission mechanism. Currently, the are no
detailed predictions on the spectra of SGR bursts.  A discussion
on the physical interpretation of our results is beyond the scope
of this \emph{Letter} and we defer it to a future work.

\newpage

\begin{acknowledgements}
We would like to thank Enrico Costa and Dario Trevese for useful
discussions, and the referee (Dr. Patrick Slane) for useful
comments.
\end{acknowledgements}

\section{REFERENCES}
\vspace{2mm}
\begin{itemize}
\setlength{\itemindent}{-8mm}
\setlength{\itemsep}{-1mm}

\item[]
Aptekar, R., et al. 2001, ApJS, 137, 227

\item[]
Boella, G., et al. 1997, A$\&$AS, 122, 327

\item[]
Cline, T., Mazets, E. and Golenetskii, S. 1998, IAU Circ.
7002

\item[] Fenimore, E.E., Laros, J.G., and Ulmer, A., 1994, ApJ,
432, 742

\item[]
Feroci, M., Frontera, F., Costa, E. et al. 1999, ApJ, 515,
L9

\item[]
Feroci, M., Hurley, K., Duncan, R.C., and Thompson, C.,
2001, ApJ, 549, 1021

\item[]
Feroci, M., Mereghetti, S., Woods, P. et al. 2003, ApJ,
596, 470

\item[] Fiore, F., Guainazzi, M., and Grandi, P., 1999, "Cookbook
for BeppoSAX NFI Spectral Analysis", available at
"http://www.asdc.asi.it"

\item[] Guidorzi, C., et al., 2004, A\&A, 416, 297

\item[]
Hurley, K., et al. 1996, ApJ, 463, L13

\item[]
Hurley, K., et al. 1999a, ApJ, 510, L107

\item[]
Hurley, K., Cline, T., Mazets, E. et al. 1999b, Nature,
397, 41

\item[]
Hurley, K., et al. 1999c, ApJ, 510, L111

\item[]
Kouveliotou, C. et al. 1993, Nature, 362, 728

\item[]
Kouveliotou, C. et al. 1998, IAU Circ. 6929

\item[]
Kouveliotou, C., Strohmayer, T.E., Hurley, K. et al. 1999,
ApJ, 510, L115

\item[]
Kouveliotou, C. et al. 2001, ApJ, 558, L47

\item[]
Laros, J.G., et al. 1986, Nature, 322, 152

\item[] Lenters, G.T. et al. 2003, ApJ, 587, 761

\item[]
Mazets, E.P., Golenetskii, S.V., Guriyan Yu.A. 1979, Sov.
Astron. Lett., 5(6), 343

\item[]
Mazets, E.P., et al . 1999, Astron. Lett. 25(10), 635

\item[]
Murakami, T., et al. 1999, ApJ, 510, L119

\item[]
Olive, J-F., et al., 2003, in Proc. "Gamma Ray Bursts and
Afterglow Astronomy 2001", G.R. Ricker and R.K. Vanderspeck eds.,
AIP 662, p. 82.

\item[] Strohmayer, T.E., and Ibrahim, A., 1998, in Proc. "4th
Huntsville Symp. on Gamma Ray Bursts", C.A. Meegan, R.D. Preece
and T.M. Koshut eds., AIP 428, p. 947.

\item[] Vasisht, G., Kulkarni, S.R., Frail, D.A., and Greiner, J.,
1994, ApJ, 431, L35

\item[]
Woods, P.M., et al., 1999, ApJL, 518, L103

\item[] Woods, P.M., Kouveliotou, C., G\"og\"us, E. et al., 2003,
ApJ, 596, 464

\item[] Woods, P.M., et al., 2004, ApJ, 605, 378
\end{itemize}

%\newpage
\clearpage

\begin{deluxetable}{lcccccccc}
 \tabletypesize{\scriptsize} \tablecolumns{9}
\tablecaption{Main characteristics of the ten selected bursts.}
\tablehead{
\colhead{Burst} & \colhead{Time } & \colhead{LECS}
& \colhead{MECS}                            & \colhead{PDS}
& \colhead{T$_{90}$}     & \colhead{Integration}      & \colhead{Peak
Flux\tablenotemark{c}}  & \colhead{Fluence\tablenotemark{c}}\\
                & \colhead{ UT }  & \colhead{Counts}
& \colhead{Counts}                          & \colhead{Counts}
& \colhead{Duration\tablenotemark{d}} & \colhead{Time}  &
\colhead{13-100 keV} & \colhead{2-100 keV}\\
                & \colhead{ SOD\tablenotemark{a} } &
\colhead{(Total/Net\tablenotemark{b})}   & \colhead{(Total/Net\tablenotemark{b})}    &
\colhead{(Total/Net\tablenotemark{b})} & \colhead{ (s) }     & \colhead{
(ms) }      & \colhead{ }   &
\colhead{  }      \\ }
\startdata
 A              &  60328        & 20               & 153/151          &
289/270          & 0.67                & 199.6               & 55.60
&  5.13 \\
 B              &  65801        & 73/36            & $>$611/469       &
$>$1008/949      & 0.23                & 200.4               & 81.25
& $>$17.95 \\
 C              &  82558        & 14               & 119/107          &
208/185          & 1.16                & 430.8               & 21.65
&  3.32 \\
 D              &  91223        & N/A              & 293/284          &
621/607          & 1.45                & 682.3               & 45.71
& 10.63 \\
 E              &  95772        & N/A              & 218/209          &
384/361          & 0.78                & 499.8               & 39.80
&  6.71 \\
 F              &  95794        & N/A              & 339/337          &
707/680          & 0.14                & 250.1               & 80.11
& 12.42 \\
 G              & 102409        & N/A              & $>$359/317       &
$>$1363/782      & 0.08                & 100.0               &
$>$150.83            & $>$13.20 \\
 H              & 111159        & 17               & 175/171          &
441/409          & 0.07                & 110.8               & 85.74
&  7.05 \\
 I              & 134665        & 13               & 158/156          &
325/298          & 0.18                & 271.0               & 47.57
&  5.82 \\
 J              & 134687        & 22               & 243/239          &
394/379          & 0.25                & 381.0               & 34.40
&  7.47 \\
\enddata
\tablenotetext{(a)} {Seconds Of Days 2001 April 18 and 19.}
\tablenotetext{(b)} {Total/Net are the burst counts Before/After
the time selections performed to accumulate the energy spectra.
For the case of LECS also an energy selection must be performed,
passing from 0.1-10 keV to 0.1-4 keV. } \tablenotetext{(c)}
{Unabsorbed, obtained by the 2BB spectral fit. The Peak Flux is
integrated over 30 ms. Units of 10$^{-8}$ erg cm$^{-2}$ s$^{-1}$
and 10$^{-8}$ erg cm$^{-2}$, respectively.} \tablenotetext{(d)}
{Determined as T$_{90}$ (the time over which 90\% of the fluence
is detected) from the MECS light curve at 10 ms resolution. The
Integration Time was instead optimized with respect to the
signal-to-noise ratio in the spectrum, and can then be sometime
very different from the T$_{90}$ duration, due to the specific
shape of the light curve in both the MECS and PDS (see for example
the case of burst "D" in Fig.\ref{burst}.}
 \label{burstlist}
\end{deluxetable}

\newpage

\begin{deluxetable}{lccccccccc}
\tabletypesize{\scriptsize} \tablecolumns{10} \tablecaption{Burst
spectral parameters for the two-blackbody model.} \tablehead{
 \colhead{Burst} &\colhead{\nh\tablenotemark{a}} &\colhead{$kT_1$}
&\colhead{R$_{_1}\tablenotemark{b}$} &\colhead{$kT_2$}
&\colhead{R$_{_2}\tablenotemark{b}$} &\colhead{\chr/dof}
&\colhead{\chr} &\colhead{$L_{1}/L{_2}$}\tablenotemark{d}\\
        &\colhead{(10$^{22}$ cm$^{-2}$)} &\colhead{(keV)}
&\colhead{(km)} &\colhead{(keV)} &\colhead{(km)} &  &  (\nh=4$\times 10^{22}$ cm$^{-2}$)\tablenotemark{c} &\\
} \startdata
 A        & 2.0 & 3.2$^{+0.5}_{-0.5}$ & 13.7  & 9.3$^{+1.9}_{-1.5}$  &
1.8    & 0.95/14     & 1.15 & 0.90$\pm{0.23}$\\
 B        & 2.5 & 3.4$^{+0.3}_{-0.2}$ & 23.5  & 9.9$^{+0.9}_{-0.8}$  &
2.7    & 0.83/43     & 0.89 & 1.00$\pm{0.16}$\\
 C        & 2.1 & 2.1$^{+0.4}_{-0.5}$ & 13.5  & 7.9$^{+0.9}_{-0.8}$  &
1.5    & 0.64/9      & 0.71  & 0.83$\pm{0.24}$\\
 D        & 2.0 & 3.3$^{+0.5}_{-0.5}$ & 9.9   & 10.2$^{+1.5}_{-1.5}$ &
1.1    & 1.27/29     & 1.59 & 0.77$\pm{0.17}$\\
 E        & 2.0 & 2.7$^{+0.5}_{-0.4}$ & 12.2  & 8.6$^{+0.9}_{-0.9}$  &
1.6    & 1.52/18     & 1.80 & 0.82$\pm{0.17}$\\
 F        & 2.0 & 3.6$^{+0.3}_{-0.5}$ & 16.5  & 10.3$^{+1.1}_{-1.3}$ &
1.9    & 0.87/32     & 1.11 & 0.90$\pm{0.19}$\\
 G        & 2.0 & 3.7$^{+0.4}_{-0.3}$ & 24.4  & 8.9$^{+0.7}_{-1.0}$  &
4.2    & 0.83/31     & 1.02 & 0.83$\pm{0.14}$\\
 H        & 2.0 & 4.1$^{+0.4}_{-0.8}$ & 15.7  & 10.5$^{+2.2}_{-2.0}$ &
2.1    & 0.54/20     & 0.69 & 0.83$\pm{0.19}$\\
 I        & 2.5 & 3.3$^{+0.4}_{-0.5}$ & 11.9  & 11.0$^{+1.8}_{-1.5}$ &
1.2    & 1.04/13     & 1.09 & 0.76$\pm{0.18}$\\
 J        & 2.0 & 3.0$^{+0.4}_{-0.4}$ & 13.9  & 9.9$^{+1.3}_{-1.2}$  &
1.3    & 0.92/19     & 1.31 & 1.03$\pm{0.21}$\\
 Average Spectrum & 2.0 & 3.4$^{+0.2}_{-0.3}$ & 13.4  & 9.33$^{+0.8}_{-0.8}$ &
1.9    & 0.62/27     & 0.92 & 0.85$\pm{0.17}$\\
Joint Fit & 2.0 & 3.34$^{+0.13}_{-0.14}$ & N/A  &
9.47$^{+0.38}_{-0.39}$ &
N/A    & 1.02/255     & 1.19 & N/A\\
\enddata
\tablenotetext{(a)} {Here \nh was constrained between 2 and 3
$\times 10^{22}$ atoms cm$^{-2}$.} \tablenotetext{(b)} {Radius of
the emitting region, assuming a conventional distance to the
source of 10 kpc.} \tablenotetext{(c)} {When \nh was allowed to
vary between 4 and 5 $\times 10^{22}$ atoms cm$^{-2}$ it went to 4
in all cases, and the best-fit parameters were consistent with
those for the case at lower \nh. For this reason we only give the
\chr~ of the fits and not the values of the parameters.}
\tablenotetext{(d)}{Ratio of bolometric luminosity of the two
components, as derived from the Joint Fit (except for the Average
Spectrum).}
 \label{fit2BB}
\end{deluxetable}

\newpage

\begin{deluxetable}{lccccccccccc}
\tabletypesize{\scriptsize} \tablecolumns{12} \tablecaption{Burst
spectral parameters for the cutoff-power law.}
\tablehead{\colhead{Burst} &\colhead{\nh\tablenotemark{a}}
&\colhead{$\Gamma$} &\colhead{$E_c$} &\colhead{\chr/dof} &
 & &  &\colhead{\nh\tablenotemark{b}}&\colhead{$\Gamma$}&\colhead{$E_c$}&\colhead{\chr}
\\
  &\colhead{(10$^{22}$ cm$^{-2}$)}
& &\colhead{(keV)} &  &
 & &  &\colhead{(10$^{22}$ cm$^{-2}$)}& &\colhead{keV}&
\\
} \startdata
 A & 3.0  & 0.30$^{+0.11}_{-0.32}$     & 14.9$^{+3.0}_{-3.3}$ & 0.99/15
& & &  & 4.30 & 0.31$^{+0.39}_{-0.20}$ & 14.6$^{+4.7}_{-2.4}$ & 0.89
\\
 B & 3.0  & -0.01$^{+0.23}_{-0.09}$    & 12.1$^{+2.3}_{-0.9}$ & 1.60/44
& & &  & 5.0  & 0.28$^{+0.26}_{-0.10}$ & 14.1$^{+1.5}_{-2.2}$ & 1.3 \\
 C & 3.0  & 0.44$^{+0.25}_{-0.22}$     & 16.2$^{+3.8}_{-4.0}$ & 1.05/10
& & &  & 4.0  & 0.43$^{+0.43}_{-0.16}$ & 15.9$^{+6.4}_{-2.9}$ & 1.06
\\
 D & 3.0  & 0.38$^{+0.19}_{-0.23}$     & 16.8$^{+3.0}_{-2.0}$ & 0.97/30
& & &  & 4.08 & 0.51$^{+0.09}_{-0.21}$ & 18.0$^{+4.6}_{-2.3}$ & 0.93
\\
 E & 2.85 & 0.25$^{+0.29}_{-0.19}$     & 14.5$^{+3.0}_{-1.0}$ & 1.44/19
& & &  & 4.0  & 0.27$^{+0.47}_{-0.13}$ & 14.4$^{+5.6}_{-1.2}$ & 1.48
\\
 F & 3.0  & 0.22$^{+0.16}_{-0.19}$     & 14.5$^{+1.8}_{-2.2}$ & 1.00/33
& & &  & 5.0  & 0.50$^{+0.09}_{-0.23}$ & 16.6$^{+2.3}_{-3.2}$ & 0.88
\\
 G & 3.0  & 0.004$^{+0.10}_{-0.23}$    & 12.2$^{+0.9}_{-2.2}$ & 1.01/32
& & &  & 4.27 & 0.04$^{+0.08}_{-0.23}$ & 12.2$^{+1.9}_{-1.3}$ & 0.94
\\
 H & 3.0  & 0.14$^{+0.09}_{-0.37}$     & 13.6$^{+1.2}_{-3.3}$ & 0.81/22
& & &  & 5.0  & 0.25$^{+0.10}_{-0.31}$ & 13.9$^{+2.6}_{-3.0}$ & 0.68
\\
 I & 3.0  & 0.43$^{+0.14}_{-0.12}$     & 19.1$^{+3.1}_{-5.9}$ & 1.77/14
& & &  & 5.0  & 0.46$^{+0.34}_{-0.15}$ & 18.3$^{+7.7}_{-2.7}$ & 1.37
\\
 J & 3.0  & 0.58$^{+0.12}_{-0.14}$     & 18.1$^{+1.9}_{-3.8}$ & 0.87/20
& & &  & 4.63 & 0.72$^{+0.19}_{-0.18}$ & 19.5$^{+3.3}_{-4.2}$ & 0.81
\\
Average Spectrum & 3.0 & 0.17$^{+0.18}_{-0.12}$ &
13.5$^{+1.7}_{-1.2}$ & 1.13/28 & & &  & 5.0  &
0.42$^{+0.12}_{-0.14}$ & 15.3$^{+1.8}_{-1.6}$ & 0.82
\\
Joint Fit & 3.0 & 0.18$^{+0.07}_{-0.06}$ & 13.9$^{+0.7}_{-0.6}$ &
1.16/265 & & &  & 5.0  & 0.40$^{+0.06}_{-0.07}$ &
15.5$^{+0.8}_{-0.8}$ & 1.04
\\
\enddata
\tablenotetext{(a)} {Here \nh was constrained between 2 and 3
$\times 10^{22}$ atoms cm$^{-2}$.} \tablenotetext{(b)} {Here \nh
was constrained between 4 and 5 $\times 10^{22}$ atoms cm$^{-2}$.}
\label{fitPLC}
\end{deluxetable}

\clearpage
%\newpage

%\end{document}

\clearpage

\newpage

\newpage
\epsscale{0.8}

\newpage

\end{document}